\documentclass[]{fairmeta}

\usepackage{xspace}
\newcommand{\bench}{E2E-SWE\xspace}

\title{E2E-SWE: Benchmarking LLMs on Building Working Codebases from Scratch}

\author[]{Hantian Ding}
\author[]{Chloe Bi}
\author[]{Jiacheng Zhu}
\author[]{John Yang}
\author[]{Matt Deitke}
\author[]{Pengcheng Yin}
\author[]{Zijian Wang}
\author[]{Rui Hou}

\affiliation{Meta Superintelligence Labs}

\abstract{Coding agents powered by large language models (LLMs) are evolving from making localized code changes to developing complete software repositories. However, evaluating repository-scale generation remains challenging: tasks must demand system-level reasoning while ensuring that all evaluated behaviors are precisely specified and independent of any particular implementation.
We introduce E2E-SWE, a benchmark for evaluating whether coding agents can build complete, functional software repositories end to end. E2E-SWE contains 186 whole-repository generation tasks spanning 11 programming languages.
Given only a natural-language specification and an empty workspace, an agent must implement a complete, installable project that satisfies a comprehensive suite of hidden tests.
Each task is constructed by a software engineer in collaboration with an LLM; together, they develop the test suite and a corresponding implementation-independent specification.
To ensure that tasks are well specified and practically solvable, we further subject them to an iterative verification process in which autonomous agents audit and repair task defects using static inspection and failures observed from real model rollouts.
Evaluating 13 frontier models, we find substantial variation in end-to-end repository generation ability, with pass@1 ranging from 11.7\% to 67.7\%, providing strong model differentiation while leaving considerable headroom for future progress.
Analysis of agent trajectories further reveals long, front-loaded reasoning patterns, highlighting the planning and system-level reasoning required to construct working codebases from scratch.}

\date{\today}
\correspondence{Hantian Ding at \email{dhantian@meta.com}}

\metadata[Code]{\url{https://github.com/facebookresearch/E2E-SWE}}

\begin{document}

\maketitle

\begin{figure}[h!]
  \centering
  \includegraphics[width=0.82\linewidth]{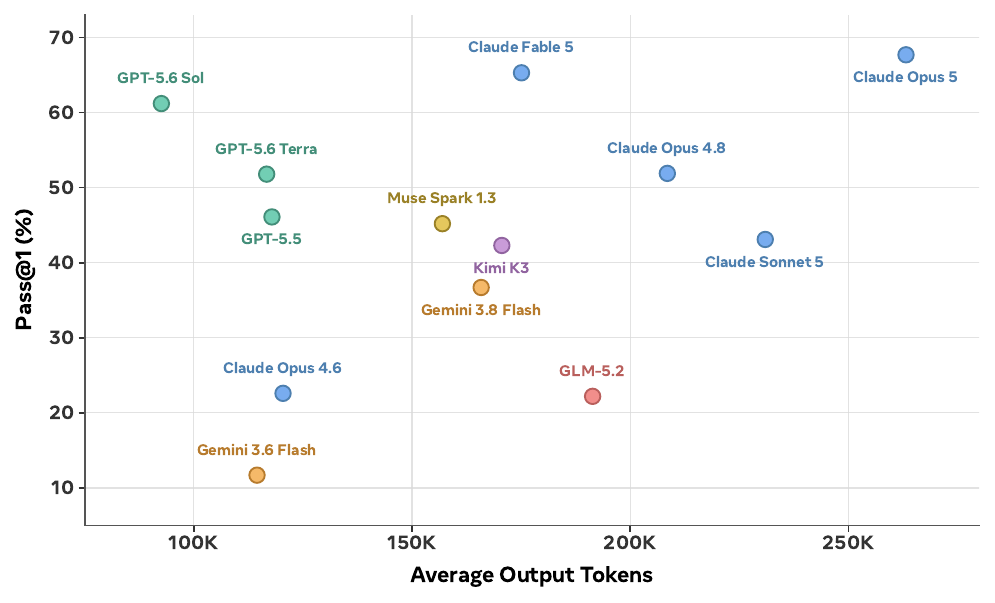}
  \caption{\bench leaderboard. Pass@1 is the average percentage of tasks fully resolved by a model.}
  \label{fig:main-results-overview}
\end{figure}

\section{Introduction}
\label{section:intro}

Large language models (LLMs) have moved to the center of modern software engineering.
In only a few years, LLM-powered tools have progressed from inline completion of individual lines and functions~\citep{humaneval,codecompose} to autonomous agents that navigate codebases and resolve real-world software issues~\citep{swebench}.
The clear trajectory is toward end-to-end software development at the repository level, where an agent is asked not for a single edit but for a complete, working system.

Creating a complete repository demands capabilities beyond the localized reasoning and editing exercised by short-horizon tasks such as function completion and issue resolution.
Whole-repository development is instead a long-horizon task requiring system-level planning and sustained execution: the model must derive a coherent design from a high-level specification, decompose requirements into interdependent components, establish consistent interfaces, and implement, integrate, and debug the system as a whole.
The challenge lies not only in implementing each component correctly, but also in making them work together while satisfying many requirements simultaneously.

To measure this capability, we introduce End2End-SWE (\bench), a multilingual benchmark for end-to-end repository generation.
Given only a natural-language specification and an empty workspace, the agent must produce a complete, installable project whose functional correctness is evaluated by executable hidden tests.
Each task is grounded in a real GitHub repository, but its specification and test suite are created from scratch by a software engineer with LLM assistance.
Together, they establish a compact behavioral contract for important user-facing functionality while leaving the internal architecture for the agent to design.
This setup mirrors a realistic two-stage workflow: a human first collaborates with an LLM to turn project goals into a concrete specification, after which a coding agent autonomously implements that specification as a working repository. \bench isolates and evaluates the second stage.

Whole-repository generation from a specification alone has been regarded as extremely challenging, with prior benchmarks~\citep{nl2repobench,beyondswe} reporting near-zero pass@1 (the fraction of tasks fully resolved in one attempt).
Yet low pass@1 meaningfully reflects model capability only when the underlying tasks are solvable.
By solvable, we mean that every behavior enforced by the hidden tests can be derived from the specification, and that any implementation satisfying the stated user-facing contract can pass regardless of how its internals are arranged.
Without this property, a failure may reflect an underspecified requirement or an overly rigid test rather than a limitation of the model.
Evidence from better-studied coding benchmarks shows this concern is substantial.
In a targeted audit of 138 SWE-bench Verified~\citep{swebench} tasks with recurring model failures, OpenAI found that 59.4\% contained material issues in test design or problem descriptions that could reject functionally correct submissions~\citep{oai_swebench_verified}.
DeepSWE separately reports that, across 789 audited SWE-Bench Pro~\citep{swebench_pro} rollouts, the benchmark's hidden tests rejected patches judged to be reasonable solutions in 24.0\% of cases~\citep{deepswe}.
These audits concern tasks that each target a single issue; a whole-repository task bundles many behavioral requirements, creating many more opportunities for a specification--test mismatch, and a single such mismatch can prevent full resolution.
Solvability is therefore especially consequential for full-repository generation.

To uphold this standard, every \bench task passes through a rigorous post-hoc review-and-repair loop.
The loop begins with static audits of the specification and tests, verifying their one-to-one correspondence.
Rollout audits then examine failures from real model attempts, distinguishing genuine implementation errors from specification gaps or overly strict tests that static inspection missed.

The final \bench dataset comprises 186 tasks across 11 programming languages, drawn from reference repositories with a median of 7.4K lines of code.
Despite the scale and complexity of these tasks, pass@1 ranges from 11.7\% to 67.7\% across 13 frontier models.
The high-end score shows that meaningful full-resolution rates are attainable when solvability is treated as a first-class design goal, while the broad spread demonstrates that pass@1 effectively distinguishes models by their ability to build complete projects from scratch.

\section{The \bench Benchmark}
\label{sec:benchmark}

\begin{figure}[t]
  \centering
  \includegraphics[width=\linewidth]{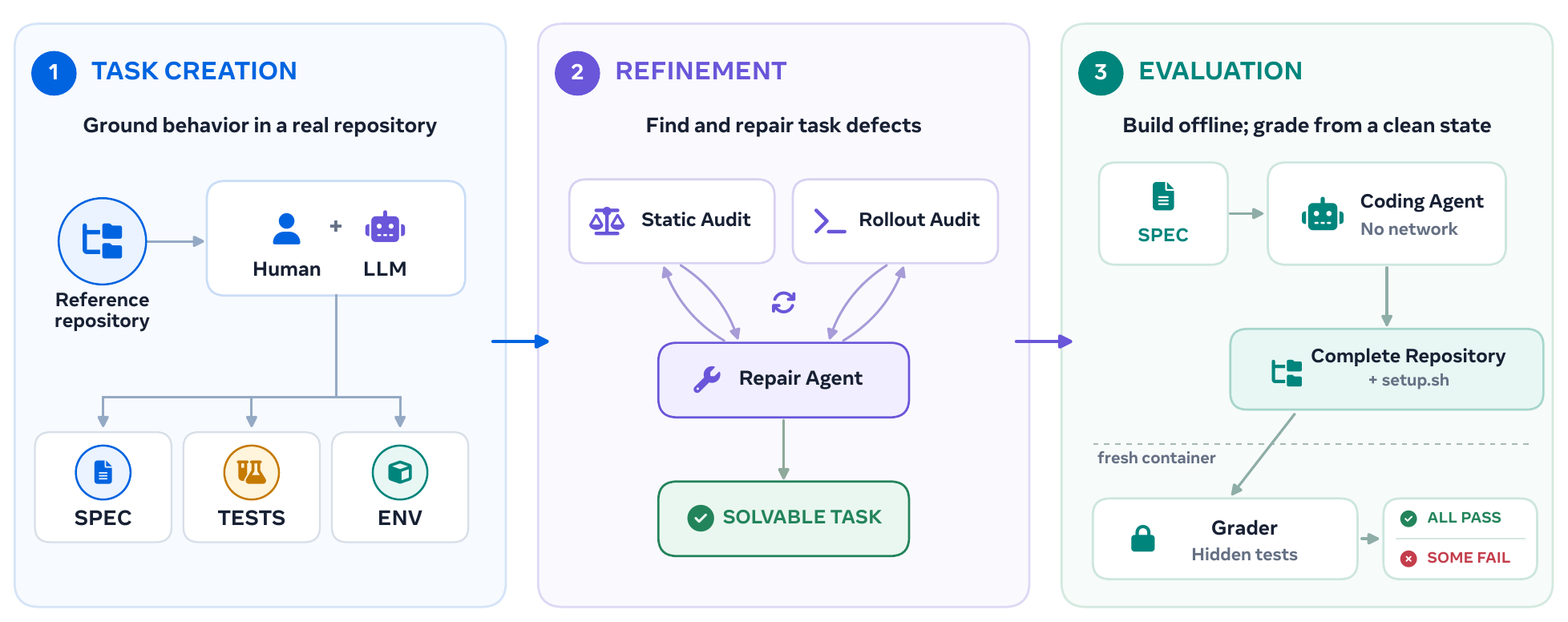}
  \caption{The \bench task lifecycle. A task author collaborates with an LLM to create a specification, hidden functional test suite, and dependency image from a reference GitHub repository. The task then passes through iterative static and rollout-guided audits, with a repair agent addressing the identified defects. During evaluation, a coding agent receives only the specification and builds a repository in an offline empty workspace. The generated repository is then installed in a fresh grading container and run against the hidden tests; full resolution requires every test to pass.}
  \label{fig:overview}
\end{figure}

We now present \bench in detail.
Figure~\ref{fig:overview} provides an overview of the task lifecycle, spanning task creation, refinement, and evaluation.

\subsection{Task Formulation}
\label{sec:task-formulation}

\bench adopts the Harbor task format~\citep{Harbor_Framework}.
Each task directory holds an \texttt{instruction.md} containing the natural-language specification shown to the agent, a \texttt{tests/} directory holding the hidden grading suite, a \texttt{solution/} directory referencing the source GitHub repository, and a \texttt{task.toml} declaring the container image and configuration.

Given an \bench task, the agent starts from an empty workspace with only the specification document, and must produce a complete repository from scratch together with a \texttt{setup.sh} script that installs the repository.
For grading, we copy the agent's implemented codebase into a fresh new container, run \texttt{setup.sh} to install the project, and then execute the hidden test suite.
This mirrors how a user clones and installs a repository, to ensure reproducibility.
A repository passes only when the code and \texttt{setup.sh} the agent committed reproduce the required behavior on their own, not because of transient development-time state such as build cache, scratch files, or incidental environment setup.

\textbf{Environment and networking.}
Each task has a dedicated container image with all required dependencies pre-installed.
Network access is disabled during both the agent's implementation phase and grading, preventing the agent from retrieving the target repository or related information from the internet.
The self-contained image allows both phases to run without any network-dependent setup.

\textbf{Solution.}
Each task ships a reference solution at \texttt{solution/solve.sh}.
It typically clones the original repository, checks out a pinned commit for reproducibility, and then composes the same short \texttt{setup.sh} the agent is expected to write.

\textbf{Grading interface.}
The test entrypoint \texttt{tests/test.sh} must emit a \texttt{ctrf.json} report that records the outcome of every individual test.
CTRF is a widely supported, language-agnostic test-reporting format, which lets \bench work with testing frameworks beyond \texttt{pytest} and, in principle, generalize to any programming language.

\subsection{Task Creation}
\label{sec:task-construction}

Each \bench task is authored by a software engineer with LLM assistance. The agent handles most of the heavy lifting, such as generating test suites that run to hundreds or thousands of lines and drafting the specification. The author plays a complementary role, reviewing the generated content, identifying gaps, steering quality improvements, and making the judgment calls on design choices.
Below we describe the step-by-step workflow that authors follow.

\subsubsection{Building the Test Suite}
\label{sec:test-suite}

Prior work on whole-repository generation often evaluates generated code using the reference repositories' test suites~\citep{commit0,nl2repobench,denovoswe}.
Those tests were designed to validate and maintain a particular codebase, not to assess the functional correctness of an alternative implementation.
They may therefore depend on private helpers or other structural details of the reference codebase that a correct reimplementation need not preserve.
We instead author a new suite for each task, consulting the original tests only to identify intended behaviors and edge cases.

Each test exercises a realistic user-facing scenario through public interfaces and asserts only observable behavior.
This specifies what the repository must do while leaving agents free to choose their internal decomposition, data structures, and algorithms.
Authors iteratively measure test coverage on the reference implementation to identify missing functionality, but do not seek to maximize coverage indiscriminately; functionality that cannot be tested or whose inclusion would add only trivial, shallow requirements may be excluded with explicit justification.

\subsubsection{Writing the Specification}
\label{sec:spec-writing}

Authors write the task specification only after creating the hidden test suite.
The specification is test-driven rather than an exhaustive description of the reference implementation.
It provides sufficient information for an agent to implement every behavior required by the tests.
Implementation details not pinned by the tests, such as internal decomposition, private helpers, and source-tree layout, are deliberately left unspecified.

The correspondence between specification and tests is central to task solvability.
Every tested behavior must be stated in the specification or follow from a standard domain convention, and every documented behavior must be exercised by the tests.
Authors audit this alignment assertion by assertion; when a test depends on a choice left open by the specification, we either make that choice an explicit requirement or relax the test to accept any reasonable choice.

\subsection{Quality Refinement}
\label{sec:task-refinement}

After authoring, each task enters an automated refinement loop spanning multiple rounds.
In each round, specialized LLM reviewers inspect the task from complementary perspectives, and a revision agent updates the specification or tests to address their findings.

\textbf{Static Review.}
Prior to model evaluation, we conduct two types of static review with independent review agents.
\emph{Test Quality} assesses the hidden suite against the reference repository without reading the specification, checking that it covers major user-facing functionality with substantive, nonredundant, and implementation-agnostic assertions.
\emph{Spec Fairness} audits the correspondence between specification and tests without access to the reference repository, verifying that each asserted behavior is explicitly documented or follows a standard domain convention and flagging tests that depend on choices the specification leaves open.

\textbf{Rollout-Guided Review.}
After static review, we run coding agents on each task and collect the test failures from their rollouts.
A reviewer examines each failure to determine whether the generated implementation fails to meet the stated requirements or satisfies them but is nevertheless rejected by the test.
The former is a genuine model error; the latter indicates a task defect such as an underspecified contract, an overly rigid assertion, or a flaky test.
By observing how independently generated code behaves under the tests, rollout-guided review can expose defects that are difficult to identify through static inspection alone.

\subsection{Dataset Statistics}
\label{sec:dataset-stats}

\begin{figure}[t]
    \centering
    \includegraphics[width=\linewidth]{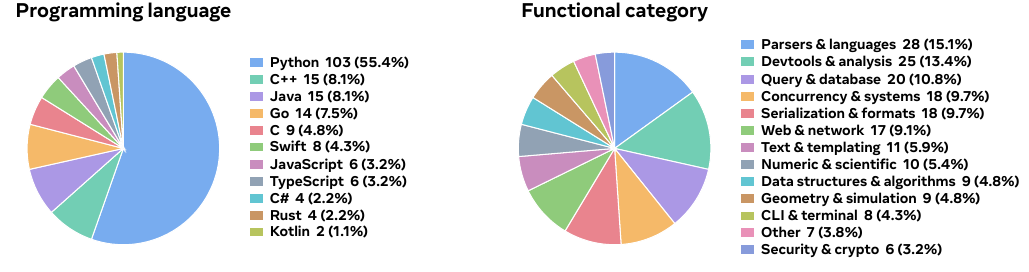}
    \caption{Composition of the \bench corpus by programming language and functional category.}
    \label{fig:dataset-composition}
\end{figure}

Our final dataset consists of 186 tasks spanning 11 programming languages and 13 functional categories.
As Figure~\ref{fig:dataset-composition} shows, \bench is diverse both in programming language and in the functionality that agents must build, ranging from parsers, developer tools, and database systems to networking, scientific computing, simulation, and cryptography.

The complexity of an \bench task stems from both the breadth of behavior expressed in its specification and the substantial implementation it requires the agent to reconstruct.
Table~\ref{tab:dataset-scale} characterizes these complementary dimensions through the scope of the authored task artifacts and the size and maturity of their reference repositories.

\begin{table}[t]
    \centering
    \begin{NiceTabular}{lrrr}
    \toprule
    \textbf{Benchmark Task} & P10 & P50 & P90 \\
    \midrule
    Hidden tests & 21 & 38 & 96 \\
    Spec (chars) & 8,460 & 16,882 & 29,206 \\
    \midrule
    \multicolumn{4}{l}{\textbf{Reference Repository}} \\
    \midrule
    Code files & 12 & 46 & 144 \\
    Code lines & 2,398 & 7,360 & 25,457 \\
    GitHub stars & 130 & 776 & 5,211 \\
    Commits & 186 & 754 & 2,512 \\
    \bottomrule
    \end{NiceTabular}
    \caption{Statistics for \bench tasks and their reference repositories, reported at the 10th, 50th, and 90th percentiles.}
    \label{tab:dataset-scale}
\end{table}

\section{Experiments}
\label{sec:experiments}

\subsection{Experimental Setup}
\label{sec:experimental-setup}

\textbf{Models.}
We evaluate thirteen frontier models across six model families: Claude, GPT, Muse Spark, Gemini, Kimi, and GLM.
We select the strongest reasoning-effort setting that runs reliably within the 128K per-request output-token limit enforced by API providers, since some models at \texttt{max} effort may exceed this limit due to excessively long reasoning in a single turn, causing API call failures.
We run each model four times independently over the full corpus and report the average.

\textbf{Agent scaffold.}
All models are evaluated with a simple ReAct agent, similar to mini-SWE-agent~\citep{sweagent} except for two additional tools: an edit tool for efficiently viewing and modifying long files and a submit tool for explicitly signaling task completion.
Each rollout has a 4-hour wall-clock budget, and is capped at 1 million context tokens and 1,000 agent turns without context compaction.
Neither the token nor agent-turn limit was reached in practice.

\textbf{Metric.}
Our primary metric is pass@1: the fraction of tasks fully resolved by a single rollout, where resolution requires the generated repository to pass every hidden test.
We do not award partial credit for passing only a subset of a task's tests.
A repository is an integrated software artifact, and one missing critical behavior can leave it unusable even when many other tests pass.
Moreover, test suites differ in their number and granularity, so fractional test scores can reflect how behavior was decomposed into assertions rather than how much of the repository was correctly reconstructed.
Full resolution instead weights each task equally and directly measures whether the model produced a complete implementation of the specified contract.

\subsection{Main Results}
\label{sec:main-results}

\begin{table}[t]
    \centering
    \small
    \begin{NiceTabular}{lcrrr}
    \toprule
    Model & Effort & Pass@1 & Agent turns & Output tokens (k) \\
    \midrule
    Claude Fable 5\textsuperscript{$\dagger$}~\citep{claude_fable_5} & high & \underline{65.3} & 56.3 & 175.1 \\
    Claude Opus 5~\citep{claude_opus_5} & xhigh & \textbf{67.7} & 120.2 & \textbf{263.2} \\
    Claude Sonnet 5~\citep{claude_sonnet_5} & xhigh & 43.1 & \underline{176.6} & \underline{231.0} \\
    Claude Opus 4.8~\citep{claude_opus_4_8} & xhigh & 51.9 & 87.6 & 208.5 \\
    Claude Opus 4.6~\citep{claude_opus_4_6} & high & 22.6 & \textbf{219.6} & 120.4 \\
    GPT-5.6 Sol~\citep{gpt_5_6} & max & 61.2 & 67.7 & 92.6 \\
    GPT-5.6 Terra~\citep{gpt_5_6} & max & 51.8 & 71.9 & 116.7 \\
    GPT-5.5~\citep{gpt_5_5} & xhigh & 46.1 & 59.5 & 117.9 \\
    Muse Spark 1.3~\citep{muse_spark_1_3} & max & 45.2 & 98.1 & 157.0 \\
    Gemini 3.8 Flash~\citep{gemini_3_8_flash} & high & 36.7 & 141.2 & 165.8 \\
    Gemini 3.6 Flash~\citep{gemini_3_6_flash} & high & 11.7 & 85.1 & 114.5 \\
    Kimi K3~\citep{kimi_k3} & high & 42.3 & 94.9 & 170.6 \\
    GLM-5.2~\citep{glm_5_2} & high & 22.2 & 137.7 & 191.4 \\
    \bottomrule
    \end{NiceTabular}
    \caption{Evaluation results on the 186-task \bench corpus, aggregated over 4 independent runs. Pass@1 is our main performance metric, measuring the percentage of tasks fully resolved. \textsuperscript{$\dagger$}Claude Fable 5 uses \texttt{high} because \texttt{xhigh} often exceeds the 128K output-token limit per request.}
    \label{tab:main-results}
\end{table}

Table~\ref{tab:main-results} reports full-repository resolution together with the average trajectory length and token usage of each model.
Claude Opus 5 achieves the highest pass@1, followed closely by Claude Fable 5, with GPT-5.6 Sol ranking third.
Even the strongest model leaves nearly one third of the tasks unresolved, indicating substantial headroom in whole-repository generation.
Meanwhile, performance is broadly distributed rather than concentrated near either a floor or a ceiling, giving \bench strong distinguishing power among current frontier models.

Models differ widely in how effectively they convert interaction and generation budgets into working software.
Opus 4.6 resolves far fewer tasks than GPT-5.6 Sol despite using more than three times as many turns.
The contrast extends to output-token usage: Fable 5 nearly matches Opus 5 while using roughly two thirds as many tokens, whereas Sonnet 5 uses almost as many tokens as Opus 5 but resolves substantially fewer tasks.

\begin{figure}[t]
  \centering
  \includegraphics[width=\linewidth]{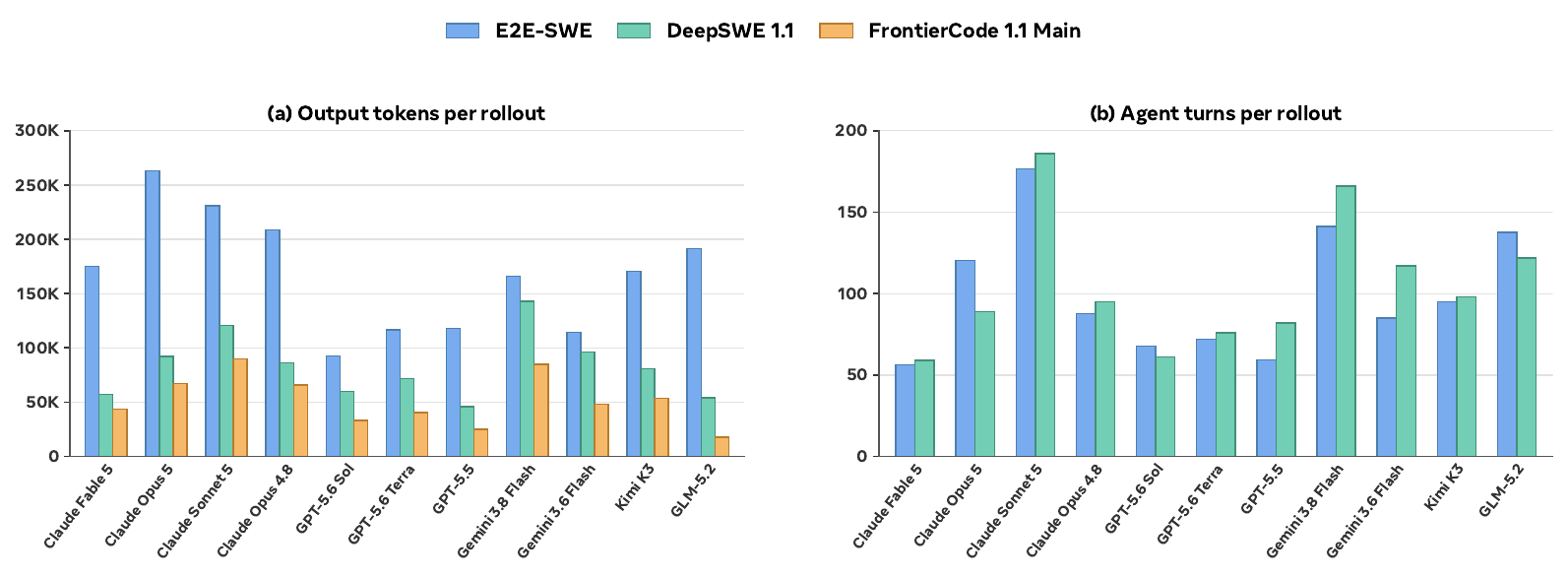}
  \caption{Rollout size across models using matched reasoning-effort settings on \bench, DeepSWE, and FrontierCode. Numbers for DeepSWE and FrontierCode are taken from their official leaderboards; FrontierCode does not report agent-turn counts.}
  \label{fig:trajectory-comparison}
\end{figure}

To characterize the complexity of full-repository generation, Figure~\ref{fig:trajectory-comparison} compares rollout size on \bench with two advanced software-engineering benchmarks.
For every matched model, \bench elicits more output tokens than DeepSWE~\citep{deepswe} and FrontierCode~\citep{frontiercode}, consistent with generating a complete repository rather than implementing a single-PR change.
Turn counts on \bench are nevertheless similar to those on DeepSWE, largely because \bench removes the initial codebase-exploration stage: it provides the complete specification upfront, whereas DeepSWE agents must navigate an existing codebase to gather context and locate the issue.

\begin{figure}[t]
  \centering
  \includegraphics[width=0.78\linewidth]{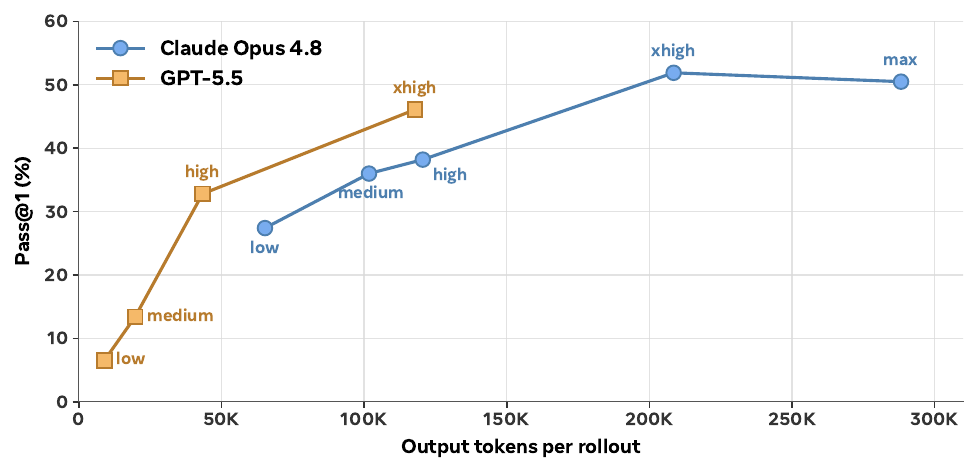}
  \caption{Inference-time scaling with reasoning effort. Each point shows pass@1 and average output-token usage at one effort setting; annotations denote effort levels and colors distinguish models.}
  \label{fig:reasoning-effort}
\end{figure}

\textbf{Impact of reasoning effort.}
We additionally sweep reasoning-effort levels for Claude Opus 4.8 and GPT-5.5, as shown in Figure~\ref{fig:reasoning-effort}.
From \texttt{low} through \texttt{xhigh}, both models use progressively more output tokens and consistently achieve higher pass@1, indicating that whole-repository generation benefits strongly from longer reasoning.
The gains eventually saturate: moving Opus 4.8 from \texttt{xhigh} to \texttt{max} consumes considerably more tokens without further improving pass@1.\footnote{Only four tasks reached the 128K output-token limit for Opus 4.8 at \texttt{max}, too few to change the overall saturation pattern.}

\section{Analysis}
\label{sec:analysis}

\subsection{Task Solvability}
\label{sec:task-solvability}

To separate signal from noise in our evaluation results, we assess whether test failures reflect genuine model limitations or flaws in the task itself.
We conduct the same rollout audit described in Section~\ref{sec:task-refinement} over one full-corpus run from each of four models: Claude Opus 5, Opus 4.6, GPT-5.6 Sol, and Kimi K3.
For each failure, an independent LLM judge (Claude Opus 5 at \texttt{max} effort) examines the specification, test, traceback, and generated implementation, then attributes the failure either to a model implementation error or to a task defect such as an underspecified requirement or over-rigid test.

Across the $186 \times 4 = 744$ model attempts, 29 (3.9\%) contain at least one failure that the judge attributes to the task.
Excluding these failures from grading would flip 16 outcomes (2.2\% of all attempts) from unresolved to resolved; the other 13 are unaffected because they also involve failures attributed to the model.

Notice that these rates should be interpreted as judge attributions rather than definitive defect counts. \bench tasks contain many interacting requirements, and the judge may overlook that a tested behavior follows from requirements distributed across the specification, thereby misclassifying a genuine model error as a task defect.

\subsection{Generated Codebases}
\label{sec:generated-codebases}

\begin{figure}[t]
  \centering
  \begin{subfigure}[t]{0.62\linewidth}
    \centering
    \includegraphics[width=\linewidth]{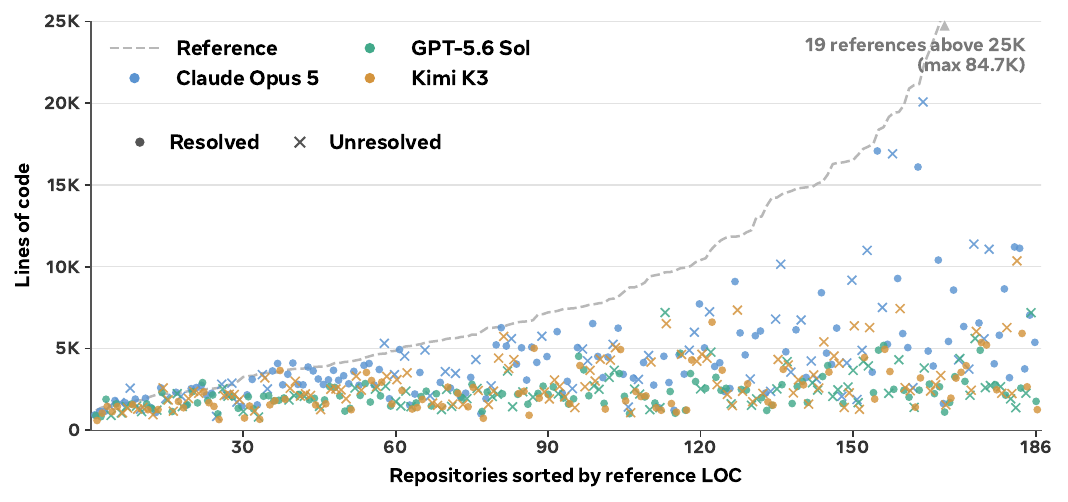}
    \caption{Per-task size against the reference.}
    \label{fig:repo-size-outcomes}
  \end{subfigure}
  \hfill
  \begin{subfigure}[t]{0.36\linewidth}
    \centering
    \includegraphics[width=\linewidth]{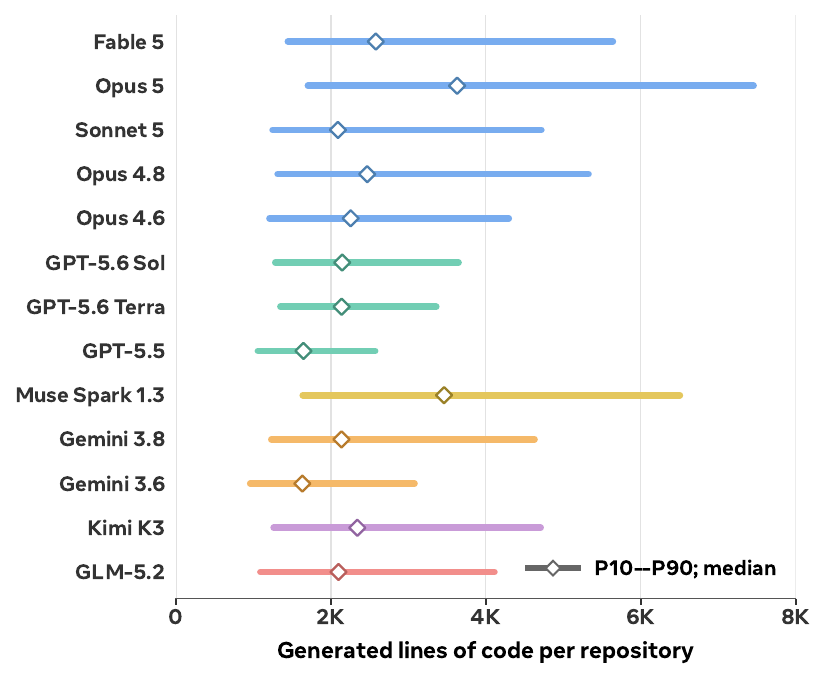}
    \caption{Size distributions across models.}
    \label{fig:repo-loc-distribution}
  \end{subfigure}
  \caption{Generated codebase sizes measured in nonblank implementation lines. \textit{Left}: repositories are ordered by reference size; colors identify models, circles denote resolved tasks, crosses denote unresolved tasks, and the reference curve is capped at 25K LOC. \textit{Right}: lines span the 10th to 90th percentiles and diamonds mark medians.}
\end{figure}

We next examine the size of the generated codebases.
Figure~\ref{fig:repo-size-outcomes} shows that models generally produce substantially fewer lines of code than the reference repositories, particularly as reference size grows. Within each model, resolved and unresolved outcomes are interspersed across the generated-size range, revealing no strong relationship between codebase size and success.

Across all models, Figure~\ref{fig:repo-loc-distribution} reveals a clear family-specific pattern. Although the Claude and GPT distributions overlap near their medians, Claude models more often produce codebases at the larger end of the range, consistent with their higher output-token usage in Table~\ref{tab:main-results}.
Generated codebase size is also largely orthogonal to model capability. Opus 4.6 and Sonnet 5, for example, produce similarly sized codebases despite their large pass@1 gap.
Together, these observations suggest that generated LOC reflects a model's coding style more than its correctness.

\subsection{Reasoning Distribution}
\label{sec:reasoning-distribution}

Table~\ref{tab:reasoning-distribution}(a) characterizes how reasoning is distributed across agent turns, reporting total reasoning together with the length, share, and position of each trajectory's longest reasoning turn.
The dominant pattern is a single, front-loaded reasoning peak.
For 11 of the 13 models, the longest reasoning turn typically occurs within the first 12 agent turns; for six of them, it occurs within the first four, even though trajectories span dozens or hundreds of turns.
These peaks can dominate a trajectory's reasoning volume: the largest reported P90 exceeds 100K tokens, and peak share exceeds 50\% in the most concentrated cases.

We attribute this shared temporal structure to the \bench task formulation.
Agents receive the complete specification upfront and have no existing codebase to explore, encouraging them to form a repository-wide plan early and spend subsequent turns implementing, testing, and revising it.

\subsection{Memorization}
\label{sec:memorization}

Each \bench task is grounded in a public GitHub repository whose source code may have appeared in the models' pretraining data.
This raises a natural question: how often does an agent solve a task by recalling the exact reference code from its memory rather than building from the given specification?
To quantify this, we propose a metric of \textit{12-gram containment}.
Within each code file, we extract overlapping 12-token windows with a stride of 1, using a model-agnostic, code-aware tokenizer.
Let $G_{12}$ and $R_{12}$ denote the resulting multisets for the generated and reference code, respectively. Containment is defined as $|G_{12} \cap R_{12}| / |G_{12}|$.
Table~\ref{tab:code-containment}(b) reports the maximum observed containment and the number of tasks exceeding 50\% and 75\% containment for each model.

\begin{table}[t]
    \centering
    \small
    \begin{tabular*}{\linewidth}{@{\extracolsep{\fill}}lrrrrrrr@{}}
    \toprule
    \multirow[c]{3}{*}{Model} & \multicolumn{4}{c}{(a) Reasoning distribution} & \multicolumn{3}{c}{(b) Code containment} \\
    \cmidrule(lr){2-5}\cmidrule(lr){6-8}
    & \multirow[c]{2}{*}{Total (k)} & Peak (k) & \multirow[c]{2}{*}{\shortstack{Peak\\Share}} & \multirow[c]{2}{*}{\shortstack{Peak\\Pos.}} & \multirow[c]{2}{*}{Max} & \multirow[c]{2}{*}{$>50\%$} & \multirow[c]{2}{*}{$>75\%$} \\
    & & P50 [P10--P90] & & & & & \\
    \midrule
    Claude Fable 5 & 123.5 & 63.0 [23.9--104.2] & 52.2\% & 2 & 98.38\% & 30 & 13 \\
    Claude Opus 5 & 180.1 & 44.0 [24.1--82.4] & 28.1\% & 4 & 97.77\% & 25 & 10 \\
    Claude Sonnet 5 & 170.3 & 28.2 [15.1--53.1] & 20.5\% & 12 & 91.87\% & 2 & 1 \\
    Claude Opus 4.8 & 158.2 & 34.5 [14.8--65.3] & 26.1\% & 8 & 98.49\% & 24 & 8 \\
    Claude Opus 4.6 & 37.5 & 2.1 [0.8--7.6] & 8.3\% & 68 & 33.34\% & 0 & 0 \\
    GPT-5.6 Sol & 47.9 & 7.3 [4.4--12.5] & 17.2\% & 3 & 14.35\% & 0 & 0 \\
    GPT-5.6 Terra & 68.5 & 9.9 [5.4--14.8] & 15.1\% & 4 & 22.60\% & 0 & 0 \\
    GPT-5.5 & 80.4 & 12.6 [9.5--16.5] & 18.5\% & 6 & 22.78\% & 0 & 0 \\
    Muse Spark 1.3 & 86.7 & 32.7 [14.7--65.7] & 43.5\% & 4 & 11.09\% & 0 & 0 \\
    Gemini 3.8 Flash & 104.8 & 21.7 [10.3--44.3] & 24.4\% & 18 & 30.50\% & 0 & 0 \\
    Gemini 3.6 Flash & 60.2 & 18.2 [7.6--32.5] & 34.2\% & 6 & 31.51\% & 0 & 0 \\
    Kimi K3 & 109.9 & 54.2 [22.7--95.2] & 55.9\% & 3 & 94.52\% & 4 & 2 \\
    GLM-5.2 & 138.5 & 26.0 [8.9--64.9] & 26.5\% & 12 & 15.83\% & 0 & 0 \\
    \bottomrule
    \end{tabular*}
    \caption{Reasoning distribution and 12-gram containment. \textbf{(a)} Reasoning is measured as output tokens excluding tool-calls. For each trajectory, Total (k) and Peak (k) are the sum and max of reasoning tokens across agent turns, respectively. Peak share is peak length divided by total reasoning, and peak position is the 1-based index of the peak turn. We report means for total reasoning and peak share, median [P10--P90] for peak length, and median for peak position. \textbf{(b)} 12-gram containment of model-generated code against the reference repository ($n=186$ tasks, one rollout per model). Max is the highest task-level containment observed for each model, and the last two columns count tasks above the indicated threshold.}
    \label{tab:reasoning-distribution}
    \label{tab:code-containment}
\end{table}

The strongest memorization signal is observed among Claude Fable 5, Opus 5, and Opus 4.8.
These models exceed 50\% containment on 24--30 tasks and 75\% on 8--13 tasks; by contrast, eight other models have no task above 50\% containment.
We further compare the grading outcomes within the $>50\%$ task group against the rest for those three models.
On the absolute scale, the resolve-rate deltas are $-0.6\,\mathrm{pp}$, $+9.6\,\mathrm{pp}$, and $+5.3\,\mathrm{pp}$ for Fable 5, Opus 5, and Opus 4.8, respectively.
Contrary to intuition, substantial reference-code overlap is far from guaranteeing resolution of the task.

\section{Related Work}
\label{sec:related-work}

Repository-level coding was initially studied in settings where the project is largely implemented.
RepoBench~\citep{repobench}, RepoCoder~\citep{repocoder}, CrossCodeEval~\citep{crosscodeeval}, and DevEval~\citep{deveval} evaluate line- or function-level completion with cross-file context.
SWE-bench~\citep{swebench} and its variants~\citep{swebench_mm,multiswebench,swebench_pro,deepswe,frontiercode}, instead target PR-level code changes to an existing repository.

Recent benchmarks push beyond localized edits toward whole-repository construction across varied task formulations and software domains.
DevBench~\citep{deveval_lifecycle} independently evaluates four stages of repository development---design, environment setup, implementation, and testing---supplying each with golden inputs rather than artifacts produced in preceding stages.
Commit0~\citep{commit0} constructs tasks from 54 Python libraries by removing function and class bodies, leaving signatures and docstrings as context for reconstructing the implementations.
RepoZero~\citep{repozero} derives 600 task instances from 35 source repositories, with each instance associated with one API-usage scenario.
Domain-specific benchmarks cover microservices in RepoGenesis~\citep{repogenesis}, interactive applications in ProjectEval~\citep{projecteval} and E2EDev~\citep{e2edev}, command-line tools in CLI-Tool-Bench~\citep{clitoolbench}, and research codebases in Paper2Code~\citep{paper2code} and PaperBench~\citep{paperbench}.
None of these benchmarks provides a domain-general evaluation of end-to-end repository creation from a project specification alone.

NL2Repo-Bench~\citep{nl2repobench} is the closest prior benchmark to \bench in both task formulation and scale. The benchmark contains 104 tasks, each asking an agent to generate a complete Python repository from scratch given only a specification document. The resulting workspace is then graded by the original repository's test suite.
In our evaluation, NL2Repo-Bench's pass@1 clusters near 10\% for all models except Claude Fable 5 and Opus 5, providing limited separation among frontier models (Appendix~\ref{app:nl2repobench-comparison}). A possible explanation is that underspecified requirements and overly strict tests make many tasks effectively unsolvable. By contrast, \bench pairs independently authored tests with test-driven specifications and iteratively reviews their behavioral alignment; empirically, it clearly separates the 13 frontier models evaluated.

ProgramBench~\citep{programbench} and MirrorCode~\citep{mirrorcode} study a complementary formulation that supplements written documentation with black-box access to a reference executable.
Their tasks jointly evaluate behavioral requirement discovery through probing and full-codebase implementation.
In some cases, determining the full set of required behaviors can be more challenging than implementing them once known, making probing the bottleneck for benchmark performance.
In contrast, \bench states the tested behaviors upfront, disentangling repository implementation from requirement discovery.
ProgramBench and MirrorCode also require the reference repository's relevant functionality to be exposed through a standalone, queryable program, limiting their applicability to repositories organized primarily around reusable APIs or other non-executable interfaces.

\section{Conclusion}

We introduced \bench, a benchmark of 186 tasks across 11 programming languages for generating complete repositories from natural-language specifications.
Its specifications and executable hidden tests are created from scratch and iteratively reviewed and repaired, making strict full-repository resolution meaningful while preserving freedom in internal design.
Across 13 models, pass@1 ranges from 11.7\% to 67.7\%, yielding clear model separation and ample headroom; a post-hoc LLM audit attributes outcome-changing failures to task issues in only 2.2\% of attempts.
Together, these results show that \bench offers a demanding yet measurable test of long-horizon software engineering capability in whole-repository generation.

\section*{Acknowledgments}

We would like to thank the following task authors for their contributions to this benchmark: Wendy Tobagus,
Daniel Molinero Reguera, Aaron Almeida, Ufuk Arslan, Yixin Bao, Chayan Bhaisare, Ben Buckley,
Jagadeesh Babu Challagundla, Raj Ray Chaudhury, Zach Coriarty, Anjan Dash, Rahul Dhruve,
German Florez, Aidan Grimshaw, Dasharath Gulvady, Jiaxing Han, Kuang Ho, Andy Huang,
Hussain Humadi, Brad Kammin, Andrei Khmylov, Noelle Li, Congxiao Lu, Ellen Lu,
Dimas Luiz Diogo de Melo Filho, Kiran Mundla, Sasha Naranjit, Rahul Rao, Shrey Shrivastava,
and Matthew Zhang.

\clearpage
\bibliographystyle{assets/plainnat}
\bibliography{paper}

\clearpage
\beginappendix

\section{Evaluation on NL2Repo-Bench}
\label{app:nl2repobench-comparison}

We evaluate all 13 models from Section~\ref{sec:main-results} on NL2Repo-Bench~\citep{nl2repobench} using the same agent scaffold and reasoning-effort settings.
To prevent agents from retrieving public reference implementations in the open-Internet setting, we disable network access and start each rollout in the grading image with all required dependencies pre-installed.

\begin{table}[h]
    \centering
    \small
    \begin{NiceTabular}{lcrrr}
    \toprule
    Model & Effort & Pass@1 & Agent turns & Output tokens (k) \\
    \midrule
    Claude Fable 5~\citep{claude_fable_5} & high & \underline{19.2} & 53.3 & 165.3 \\
    Claude Opus 5~\citep{claude_opus_5} & xhigh & \textbf{22.1} & 82.1 & \textbf{184.6} \\
    Claude Sonnet 5~\citep{claude_sonnet_5} & xhigh & 13.5 & \textbf{167.5} & \underline{171.4} \\
    Claude Opus 4.8~\citep{claude_opus_4_8} & xhigh & 13.5 & 83.8 & 158.7 \\
    Claude Opus 4.6~\citep{claude_opus_4_6} & high & 8.7 & 129.8 & 68.5 \\
    GPT-5.6 Sol~\citep{gpt_5_6} & max & 7.7 & 73.7 & 95.7 \\
    GPT-5.6 Terra~\citep{gpt_5_6} & max & 8.7 & 61.4 & 100.6 \\
    GPT-5.5~\citep{gpt_5_5} & xhigh & 12.5 & 33.9 & 74.2 \\
    Muse Spark 1.3~\citep{muse_spark_1_3} & max & 9.6 & 84.7 & 88.9 \\
    Gemini 3.8 Flash~\citep{gemini_3_8_flash} & high & 8.7 & \underline{166.6} & 134.4 \\
    Gemini 3.6 Flash~\citep{gemini_3_6_flash} & high & 8.7 & 97.6 & 87.1 \\
    Kimi K3~\citep{kimi_k3} & high & 14.4 & 78.8 & 135.4 \\
    GLM-5.2~\citep{glm_5_2} & high & 9.6 & 123.0 & 134.4 \\
    \bottomrule
    \end{NiceTabular}
    \caption{Evaluation results on the 104-task NL2Repo-Bench corpus. Each model is evaluated in one full-corpus run. Pass@1 is the percentage of tasks fully resolved; agent turns and output tokens are averages per rollout.}
    \label{tab:nl2repo-results}
\end{table}

Table~\ref{tab:nl2repo-results} shows a compressed performance range.
Except for Claude Opus 5 and Fable 5, every model resolves only 8--15 tasks, with many scores separated by just one or two tasks.
Pass@1 is lower than on \bench for every model.
However, two indicators suggest that NL2Repo-Bench is unlikely to be more complex than \bench.
The reported median reference-repository size falls between 1.5K and 4K lines of code, well below \bench's median of 7,360, and models also generate fewer output tokens per rollout on NL2Repo-Bench.
Together, these observations suggest that its low, tightly clustered scores reflect factors beyond task complexity, limiting its resolution among frontier models.

\end{document}